\documentstyle[epsf,12pt]{article}
 \def\simlt{\lower.5ex\hbox{$\; \buildrel < \over \sim \;$}}
  \def\simgt{\lower.5ex\hbox{$\; \buildrel > \over \sim \;$}}
\begin{document}

\begin{center}
\textbf{ \Large On Breaking Cosmic Degeneracy} \\
\end{center}
\begin{center}
R. Benton Metcalf and Joseph Silk \\
{\it Departments of Physics and Astronomy, and Center for Particle
Astrophysics \\ University of California, Berkeley, California 84720}
\end{center}

\begin{abstract}
It has been argued that the power spectrum of the 
anisotropies in the Cosmic Microwave Background (CMB) may be effectively 
degenerate, namely that the observable spectrum does not determine a unique 
set of cosmological parameters.  We describe the physical origin of this 
degeneracy and show that at small angular scales 
it is broken by gravitational lensing: effectively degenerate 
spectra become distinguishable at $\ell \sim 3000$ because lensing causes 
their damping tails to fall at different rates with increasing $\ell$.
This effect also helps in distinguishing nearly degenerate power spectra 
such as those of mixed dark matter models.
Forthcoming interferometer experiments should provide the means of measuring
otherwise degenerate parameters at the $5-25\%$ level.
\end{abstract}

It has recently  been  pointed out \cite{ZSS} that in a parameter space including 
open models the contours of the estimated likelihood function for planned experiments are 
highly elongated in certain directions.  This indicates that there is a near degeneracy 
between cosmological parameters which will limit the accuracy with which the individual  
parameters can be measured using only CMB observations. Note that with
sufficiently good data as expected from planned satellite experiments,
no such degeneracy arises in the more
limited subset of flat models, contrary to an earlier claim  \cite{BDS95}.  Our 
intent with this letter is to first elucidate the physical origin of this effective 
degeneracy and then show how it is broken at small angular scales.  We will not attempt 
to make detailed estimates of the precision with which particular future experiments may 
determine cosmological parameters.

The cosmic degeneracy has a simple physical explanation which can be used to easily identify 
models that have near degenerate CMB power spectra.  We  
restrict our discussion to homogeneous, isotropic cold dark matter models
with
arbitrary spatial curvature, adiabatic scalar perturbations and no tensor perturbations.  A 
curve that produces degenerate spectra can be found by first specifying $h^2 \Omega_m$, 
$h^2 \Omega_b$ and the primordial spectrum.  This fixes the comoving scale of the acoustic 
peaks.  
One then varies 
$\Omega_m$ (or h) and $\Omega_\Lambda$ in such a way that the angular size distance to the 
surface of last scattering remains constant.  In doing so,
the 
change in the redshift of the surface of last scattering must be taken into account,
\begin{equation}
z_* \cong 10^3 \Omega_b^{-0.027/(1+0.11\ln\Omega_b)}
\end{equation}
\cite{Hu95}.  The Hubble parameter is $H_o=100 h$~km/s/Mpc and $\Omega_m$, 
$\Omega_b$ and $\Omega_\Lambda$ are, respectively, the density in matter, baryons and 
vacuum energy measured in units of the critical density.  In contrast to
the apparently similar but non-degenerate behavior found for a grid of models with 
$\Omega_\Lambda + \Omega_m = 1$ \cite{BDS95}, this prescription gives models whose 
spectra are completely degenerate at small angles, or large multipole number $\ell$.  
This is what we mean by the description ``effectively degenerate''  as opposed to a 
case of ``near'' degeneracy which may arise due to experimental limitations.

\begin{figure}
\centerline{\epsfysize=3.0 in \epsffile{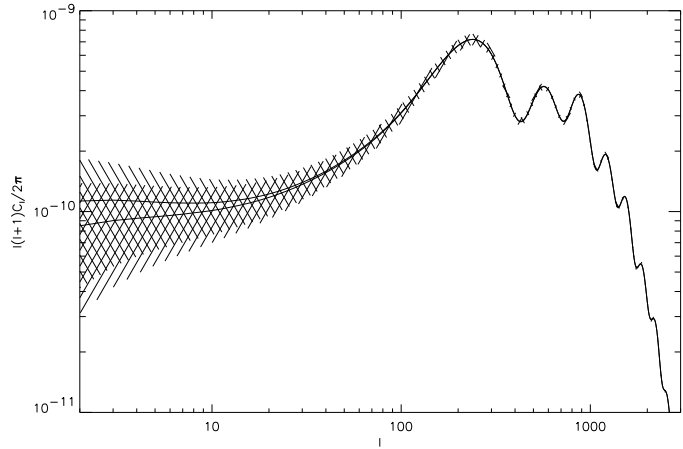}}
\caption[fig1.eps]{\footnotesize An example of effectively degenerate spectra.  For these 
models $h^2 \Omega_b=0.015$ and $h^2 \Omega_m=0.15$.  For the top model $\Omega_m=0.31$ and 
$\Omega_\Lambda=0.61$.  For the bottom one $\Omega_m=0.60$ and $\Omega_\Lambda=0.34$.  
The $1 \sigma$ cosmic variance for each of the models is shown with the hash lines 
assuming full-sky coverage.}
\label{degen} 
\end{figure}

 However, as can be seen in Figure~(\ref{degen}) these spectra are not degenerate at 
large angular scales.  This difference is due to the decay in the potential fluctuations 
in $\Omega_m<1$ models, the Integrated Sachs-Wolfe (ISW) effect. 
 Unfortunately this difference 
is of limited use for discriminating between models because of the magnitude
of  the cosmic 
variance  on these scales.  The cosmic variance is the unavoidable uncertainty that 
results from comparing a cosmological model which predicts only the statistical 
distribution of observables 
with a finite sampling of that distribution as represented by the data: there are 
many possible skies but we observe only one.  If the entire sky could be used for measuring 
the CMB anisotropies, the variance in the power spectrum would be $\sigma^2_{C_\ell}
= 2\langle C_\ell \rangle^2/(2\ell +1)$ where gaussianity has been assumed.  In addition, 
cutting out parts of the sky, such as  the galactic plane, will increase this by a factor
that is  approximately inversely proportional to the fraction  
of the sky sampled.  The cosmic variance is indicated in Figure~(\ref{degen}) by the 
hashed lines.  Note that if the parameter space is increased to include models with 
tensor perturbations, a slope nearly equivalent to the ISW effect can be
generated, making the spectra even more degenerate.
It is evident that cosmic variance prevents us from differentiating between models 
whose $C_\ell$ spectra differ only on large angular scales.

  In summary, if one set of parameters is found to fit the observed CMB power 
spectrum, a whole family of models can be found with the above prescription which fits the 
spectrum equally well except in the small $\ell$ region where the 
cosmic variance is  largest.  One could always hope to use other observables such 
as supernova surveys and more direct measures of the Hubble parameter to break 
the degeneracy in the CMB power spectrum, as suggested by
Zaldarriaga, Spergel \& Seljak \cite{ZSS}.  This is a promising approach, but is
unlikely to provide the accuracy of parameter estimation 
that has been claimed for the CMB fluctuations.  In this Letter we show that in standard 
models the observable CMB power spectrum is in fact not degenerate on small scales due to 
the processing of the background light after it leaves 
the surface of last scattering.  Gravitational lensing has the effect of both smoothing 
out the acoustic peaks and troughs in the power spectrum (\cite{seljak} and references 
therein) and of making the damping tail a less steep function of $\ell$ 
\cite{MS,BS87}.  It is the latter effect that we concentrate on here.  

Gravitational lensing by large-scale structure (LSS) magnifies and demagnifies patches of the 
sky with very nearly the same probability.  However,  the steep damping 
of structure with decreasing scale on the surface of last scattering results in a bias in 
the lensing which transfers power from lower to higher $\ell$ in the damping tail
of the power spectrum \cite{MS}.  
This results in the power spectrum falling less rapidly with 
increasing $\ell$.  Since the lensing contribution is different for otherwise degenerate 
spectra, the degeneracy will be broken and the fractional difference between two such 
models will increase with increasing $\ell$.  The strength of this lensing effect 
increases with increasing $\Omega_m$ and is generally greater the more steeply the damping tail drops.

The transformation of the $C_\ell$'s by lensing is given by 
\begin{eqnarray}
C_\ell^{ob} & \simeq & \sum_{\ell'=0}^{\infty} C_{\ell'} \frac{2\ell'+1}{2} \int_0^{\pi} ds 
\sin(s) P_\ell[\cos(s)]\left\{ e^{-\ell'^2\langle \beta_{\parallel}(s)^2 \rangle/2} P_{\ell'}[x] \right.
\label{Ctrans} \\
& & \left. +\frac{1}{2}\left[ \langle \beta_{\parallel}(s)^2 \rangle - \langle \beta_\perp(s)^2 \rangle \right] P'_{\ell'}[x] 
\right\}_{x=\cos(s)} \nonumber
\end{eqnarray}
where $C_\ell^{ob}$ is the observed power spectrum and the second moments of the components of 
the relative deflection of two light paths are given by
\begin{equation}
\left\langle \beta_{\parallel,\perp}(s)^2 \right\rangle  \simeq 
\left( \frac{2}{g(r)} \right)^2 \int_0^r d\overline{r} \int \frac{dk}{2 \pi} k^3 
P_{\phi}(k,{\overline\tau})  g(r-\overline{r})^2 \left\{ 1- J_o[k s g(\overline{r})] 
\pm J_1[k s g(\overline{r})]/k s g(\overline{r}) \right\}. 
\label{beta2}
\end{equation}
where $g(r)=\{ R\sinh(r/R),r,R\sin(r/R) \}$ for the open, flat and closed models respectively.  
The curvature scale is $R=|H_o \sqrt{1-\Omega-\Omega_{\Lambda}}|^{-1}$, 
$P_{\phi}(k,{\overline\tau})$ is the power spectrum of potential fluctuations and $r$ is the 
coordinate distance to the surface of last scattering.  This result is 
derived elsewhere  \cite{MS}.  An effectively numerically equivalent expression was obtained by 
Seljak \cite{seljak}.

\begin{figure}
\centerline{\epsfysize=9.0 cm \epsffile{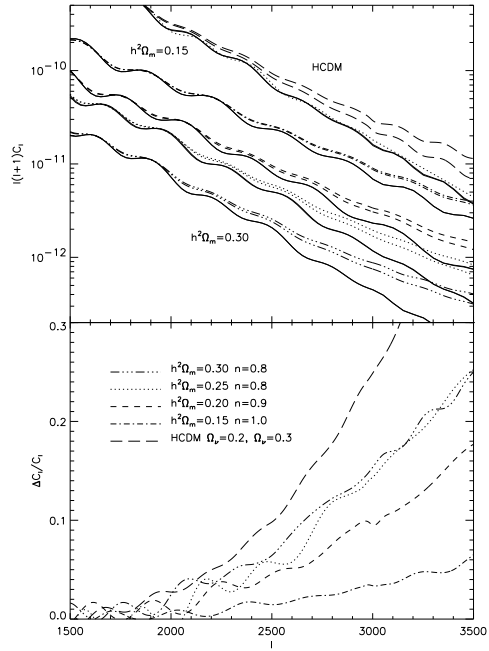}}
\caption[fig2t.eps]{\footnotesize The damping tails of CMB power spectra.  For the lower four sets 
of spectra in the top panel the unlensed spectra (all of them) are shown in solid 
lines and the lensed spectra are dotted, dashed etc.  The upper 
spectra of each of these sets have $h=0.7$ and the lower have $h=0.5$ except in the 
case of $h^2 \Omega_m=0.30$ which has $h=0.55$ and $0.7$.  The top set of spectra 
labeled HCDM are for mixed dark matter models except the dotted spectrum which is 
unlensed CDM with $\Omega_m=1$, $h=0.6$.  The normalizations have been changed for 
display purposes.  The bottom panel is the fractional difference between the lensed 
spectra.}
\label{fig2}
\end{figure}

Figure~(\ref{fig2}) shows four sets of effectively degenerate spectra and two mixed 
dark matter spectra which will be discussed later.  For all of the models, $h^2 
\Omega_b=0.015$ and each set has a different value of  $h^2 \Omega_m$.  The matter power spectrum 
used in these calculations is of the form
 $P(k)= A^2 k^n T^2(k/h\Gamma)$. 
 We use the standard CDM transfer function of \cite{bard86}
with $\Gamma=\Omega_m h \exp(-\Omega_b-\Omega_b/\Omega_m)$ 
\cite{sugy95}.  For each set of equal $h^2 \Omega_m ,$ the 
matter power spectrum in the model with 
the lowest $h$ is COBE-normalized  \cite{bunn97}.  
The conjugate  model of each set is normalized so that its CMB power spectrum is degenerate with the 
COBE-normalized spectrum.  This requires changing the normalization by a factor of $\sim 1.14-1.02$ 
with the lower range being for the higher $h^2 \Omega_m$.
 The nonlinear evolution of the matter power spectrum is followed using the 
method of Peacock \& Dodds \cite{peac96}.  The nonlinear evolution of the power spectrum contributes 
significantly to lensing at these angular scales.  All of the intrinsic, pre-lensed CMB 
power spectra are calculated using the Seljak \& Zaldarriaga \cite{seljak96} computer 
code up to $\ell=3000$.  Beyond this point we use an extrapolation of the spectrum.

The bottom panel of figure~(\ref{fig2}) gives the fractional difference between the
spectra that would be degenerate were there no lensing.  The splitting of the spectra is 
more pronounced as $h^2 \Omega_m$ gets larger.  The normalizations of the spectra in the 
top panel have been changed for display purposes (otherwise the lensed spectra for 
the $h^2 \Omega_m = 0.15-0.30$ models would overlap).  All matter power spectra are COBE-normalized and satisfy the $\sigma_8$ constraint on the variance in the average overdensity in a $8 h^{-1}$Mpc sphere
as determined from the abundance of galaxy clusters \cite{VL96}.
The low $h$ model of the $h^2 \Omega_m=0.30$ 
models has $h=0.55$ instead of $h=0.5$ so that $\Omega_m \leq 1$ for all the models.  The 
$h^2 \Omega_m =$ $0.20$, $0.25$ and $0.30$ models have tilted primordial spectra so that 
their $\sigma_8$'s satisfy the  
cluster abundance constraints.  The reduction 
in the strength of the lensing in the tilted models is partially made up for by the 
steepening of the CMB power spectrum which acts to increase the enhancement.  Most of 
the difference in the lensed models is due to the difference in the implied COBE normalization. 
 The differences in the global geometry and the shape and 
evolution of the power spectrum play a smaller role.  The parameters for each of the models are 
given in Table~\ref{table1}.

\begin{table}[t]
\caption{Model Parameters \label{table1} }
\begin{center}
\begin{tabular}{cccccc}
 $h^2 \Omega_m$ & $h^2 \Omega_b$ & $h$    & $\Omega_\Lambda$ & $n$   & $\sigma_8$ \\ \hline \hline

  0.30         & 0.015        & 0.55 & 0.003          & 0.8 & 0.85     \\ \hline
  0.30         & 0.015        & 0.70 & 0.34           & 0.8 & 0.88     \\ \hline
  0.25         & 0.015        & 0.50 & 0.00           & 0.8 & 0.78     \\ \hline
  0.25         & 0.015        & 0.70 & 0.43           & 0.8 & 0.81     \\ \hline
  0.20         & 0.015        & 0.50 & 0.17           & 0.9 & 0.84     \\ \hline
  0.20         & 0.015        & 0.70 & 0.52           & 0.9 & 0.90     \\ \hline
  0.15         & 0.015        & 0.50 & 0.17           & 1.0 & 0.84     \\ \hline
  0.15         & 0.015        & 0.70 & 0.52           & 1.0 & 0.90     \\ \hline
 \end{tabular}
\end{center}
\end{table}

Adding more parameters  of course complicates things, but lensing of the damping tail is 
likely to remain a strong discriminator between models.  Tensor modes contribute 
only to the large scale anisotropies where lensing has no affect so they cannot produce 
an additional set of effectively degenerate models, but
only add extra dimensions to the degenerate 
parameter subspace.  If the tensor mode contribution is large enough, it can be seen 
over cosmic variance ($T/S \simgt 0.14$ \cite{KT94}), otherwise it might be detectable 
in the polarization anisotropies.  Adding hot dark matter changes the heights and 
positions of the acoustic peaks at intermediate scales, but if there is only one 
species of neutrino with a cosmologically interesting mass ($0.4 \simgt \Omega_{\nu} 
\simgt 0.2
, m_{\nu}=46.6\Omega_{\nu}h^2$eV) these changes are quite small and will be difficult 
to detect even with future satellite missions \cite{DGS}.  There are two mixed dark 
matter (HCDM) spectra in Figure~(\ref{fig2}), both with $\Omega_m+\Omega_{\nu}=1$ and 
$h=0.6$.  The lensing for these models was calculated with only the linear evolution of 
the matter power spectra given by \cite{PS95}.  Nonlinear evolution would probably 
increase the difference between the lensed spectra.  It is already evident that lensing 
of the damping tail would be a useful probe of hot dark matter if most of the other 
cosmological parameters were to some degree constrained by other observations.  There 
remains the possibility that additional effective or near degeneracies exist in this or 
other expanded parameter spaces.  This will always be a possibility because of the 
unlimited number of additional parameters that could be added.

Radio interferometry presently holds the most promise for observing the CMB at the small 
angular scales that are necessary to observe these lensing 
effects.  The Cosmic Background Interferometer (CBI) is now being built
\cite{Carlstrom}.  This is an 
array of $13$ one meter dishes that is expected to measure the CMB power spectrum in 
the range $690 < \ell < 3800$ at $27-36$GHz.  The width of the $\ell$-space window function, 
$\Delta\ell$, for an interferometer is $\sim D$, the diameter of the dishes in units of 
wavelength.  The width may be further reduced by mosaicing, and
 $\Delta\ell \sim 50-100$ is 
expected.  Widths of this order do not significantly change the predicted difference between 
lensed spectra.  The sample variance is given by $\sigma^2_{sam} \simeq (4\pi/A) \sigma^2_{cos}
\Delta\ell$ where $A$ is the solid angle covered.  For an instrument such as
 CBI, reducing the 
sample variance  to $10\%$ at $\ell=3000$ will require covering an area of $13-27 
\mbox{ deg}^2$ or for $20\%$ accuracy $3-7 \mbox{ deg}^2$, depending on $\Delta\ell$.

A possible complication to these lensing predictions is a wide class of foregrounds that 
may be present at these angular scales.  Among these 
are radio galaxies, the kinematic and thermal Sunyaev-Zel'dovich effect from galaxy 
clusters and perhaps the most worrying, the kinematic Sunyaev-Zel'dovich effect from 
either bubbles of ionized gas surrounding quasars \cite{ADPG} or Ly$\alpha$ absorption 
systems \cite{L96}.  These effects adds a component to 
the power spectrum at small angular scales without distorting the thermal spectrum.  This 
decreases the steepness of the damping tail which reduces the lensing effect.  On the 
other hand if reionization is uniform it will damp the spectrum and increase the effect 
of lensing.  The majority of the lensing takes place after reionization affects the CMB.  
The Rees-Sciama effect is not likely to change our predictions significantly because it 
is expected to contribute only $\sim 1\%$ or smaller to the power spectrum 
at the scales of interest \cite{seljakb}.  The extent to which the Vishniac effect 
contributes to the spectrum is strongly dependent on the ionization history of the 
universe, but is probably small \cite{HSS,DJ}.  In any case, the contribution 
to the CMB power spectrum from these two effects will generally increase with increasing 
small-scale structure just as lensing does, 
and will then complement lensing in splitting the degenerate spectra.  Foregrounds may 
lift the degeneracy of spectra by themselves, but to determine in what way this might 
happen it is necessary to relate the ionization history of the universe to the underlying 
cosmological model.  At present this connection is very uncertain.

In summary, we have shown that if sufficient sensitivity and low enough sample variance 
can be obtained with future radio interferometers, the problem of degeneracy in the CMB 
power spectrum can be reduced and effectively removed in many models.  Lensing can 
separate degenerate spectra on the $\sim 5-25\%$ level over a wide range of $\ell$ space.
 Because the range is wide, distinguishing between models would require significantly 
less precision in individual $C_\ell$'s.  It is the rate at which the damping tail falls 
that distinguishes models.  Removing the degeneracy would allow for more accurate 
measurements of the cosmological parameters based on CMB measurements alone and would 
enable us to differentiate more definitively between the pre- and post-recombination 
states of the universe.

\leftline{}
We would like to thank the referee for helpful comments and suggestions.  This research has been supported in part by grants from NASA and DOE.

\end{document}